\documentclass[prb,showpacs,twocolumn]{revtex4}

\newcommand {\LSVO}{La$_{1-x}$Sr$_x$VO$_{3}$}
\newcommand {\NaxCo}{Na$_x$CoO$_2$}
\newcommand {\EparaC}{E $\parallel c$}
\usepackage{graphicx}
\usepackage{bm}
\usepackage{pifont}
\usepackage{amsmath}
\begin{document}
\title{Thermoelectric response in the incoherent transport region near Mott transition: the case study of {\LSVO}}
\author{M. Uchida$^1$, K. Oishi$^1$, M. Matsuo$^{2,3}$, W. Koshibae$^4$, Y. Onose$^{1,5}$, M. Mori$^{2,3}$, J. Fujioka$^5$, S. Miyasaka$^6$, S. Maekawa$^{2,3}$, and Y. Tokura$^{1,4,5}$
} 
\affiliation{$^1$Department of Applied Physics, University of Tokyo, Tokyo 113-8656, Japan\\$^2$Advanced Science Research Center, Japan Atomic Energy Agency (JAEA), Tokai, Ibaraki 319-1195, Japan\\ $^3$CREST, Japan Science and Technology Agency (JST), Tokyo 102-0075, Japan\\$^4$Cross-Correlated Materials Research Group (CMRG) and Correlated Electron Research Group (CERG), ASI, RIKEN, Wako 351-0198, Japan\\$^5$Multiferroics Project, ERATO, Japan Science and Technology Agency (JST), Tokyo 113-8656, Japan\\$^6$Department of Physics, Osaka University, 1-1 Machikaneyama, Toyonaka, Osaka 560-0043, Japan}

\begin{abstract}
We report a systematic investigation on the high-temperature thermoelectric response
in a typical filling-control Mott transition system {\LSVO}.
In the vicinity of the Mott transition, incoherent charge transport appears with increasing temperature
and the thermopower undergoes two essential crossovers, asymptotically approaching the limit values
expected from the entropy consideration, as known as Heikes formula.
By comparison with the results of the dynamical mean field theory, 
we show that the thermopower in the Mott critical state
mainly measures the entropy per charge carrier that depends on electronic degrees of freedom available at the measurement temperature.
Our findings verify that the Heikes formula is indeed applicable to the real correlated electron systems
at practical temperatures ($T>200$ K).
\end{abstract}
\maketitle

\section{Introduction}

Thermoelectric materials, which interconvert temperature gradients and electric voltages,
could play a key role for future energy production and utilization.\cite{review1}
While the conventional research and optimization of the thermoelectric materials
have been done based on the one-electron band theory,
transition-metal oxides with strong electron correlation have recently attracted much attention as promising candidates
since the discovery of unusually large thermopower $S$ in {\NaxCo}.\cite{NaxCo0, NaxCo1}
One possible origin of the enhanced $S$ in the correlated metals is
entropy flow of the correlated electrons,\cite{koshimae1,koshimae2,koshimae3,NaxCo5,kotliar}
expressed as Heikes formula.\cite{heikes1, heikes2}
This mechanism could provide a novel strategy
for designing the high-performance thermoelectric materials in a different way from the band-theory approach.
However, the Heikes formula is strictly correct only in the high-temperature limit,\cite{heikes1, heikes2}
and its applicability at practical temperatures is highly nontrivial.
It has been suggested by some groups that the large $S$ observed in {\NaxCo}
can be explained within some band structure effects,\cite{NaxCo2,NaxCo3,NaxCo4}
and the applicability of the Heikes formula to real materials is still under debate.

As derived by the Boltzmann transport equation in the band theory,
conventional metals show the very small $S$ well below the Fermi temperature $T_{\mathrm{F}} \sim 10^4$ K, 
because it is strongly reduced by the factor $T/T_{\mathrm{F}}$ from the value $k_{\mathrm{B}}/e \simeq 86$ $\mu \mathrm{V}/\mathrm{K}$.
In the high-temperature limit, on the other hand, 
the conducting state is treated as the incoherent hopping of localized charge carriers between the adjacent sites,
and hence $S$ derived by the Heikes formula becomes of the order of $k_{\mathrm{B}}/e$ reflecting the entropy consideration.
In materials with strong electron correlation, 
there exist two possible cases for the high-temperature regime ($k_{\mathrm{B}}T \gg W$):    
(i) $k_{\mathrm{B}}T \ll U$ or (ii) $U \ll k_{\mathrm{B}}T$,
where $W$ is the half bandwidth, $U$ the on-site Coulomb interaction.
In the case of (i) $k_{\mathrm{B}}T \ll U$, $S$ obtained from the Heikes formula is expressed as
\begin{equation}
\begin{split}
S_1=-\frac{k_{\mathrm{B}}}{e}\ln \frac{g_e}{g_h}-\frac{k_{\mathrm{B}}}{e}\ln \frac{x}{1-x},
\end{split}
\end{equation}
where $x$ is the hole concentration,
and $g_e$ ($g_h$) denotes the local degeneracy of the electronic configuration on the site without (with) hole carrier.\cite{koshimae1,koshimae2,koshimae3}
In this temperature region, the spin and orbital degrees of freedom explicitly contribute to $S$.
For (ii) $U \ll k_{\mathrm{B}}T$, the carrier hopping occurs, allowing the doubly occupied states, and $S$ is given by
\begin{equation}
\begin{split}
S_2=-\frac{k_{\mathrm{B}}}{e}\ln \frac{2m-n}{n},
\end{split}
\end{equation}
where $m$ and $n$ are the orbital degeneracy and the electron number per site, respectively.\cite{heikes1,heikes2}
Therefore, it is predicted that high-temperature $S$ in the correlated metals shows crossovers
asymptotically approaching the large limit values $S_1$ and $S_2$.
In addition, depending on the signs of $S_1$ and $S_2$,
the temperature dependence of $S$ could be non-monotonic as accompanied by nontrivial sign changes,
while the sign of $S$ in the conventional metals is determined by the sign of charge carrier.
It is expected that by observing such crossovers of $S$,
the applicability of the Heikes formula can be experimentally verified.

As a model system suitable for examining the Heikes formula,
we chose a canonical filling-control insulator-metal (Mott) transition system {\LSVO},\cite{mott,MIT,lsvo1}
where a metallic state with coherent charge transport realizes only at low temperatures
and the incoherent charge transport appears above $\sim 200$ K centered around the critical doping level $x_{\mathrm{IM}}$.
Thus, around $x_{\mathrm{IM}}$, $S$ is anticipated to asymptotically approach the Heikes-formula values with increasing temperature
in a manner dependent on the doping level and the Coulomb correlation.
By measuring $S$ up to 1250 K and comparing to the results of dynamical mean field theory (DMFT) for the Hubbard model,
we have clarified the essential features of the correlation effect on the high-temperature $S$
and verified the applicability of the Heikes formula to the real correlated electron systems.

\section{Experimental and calculation procedures}

Single-crystalline ($x=0$-0.26) and polycrystalline ($x=0$-0.80) samples of {\LSVO} were prepared
by a floating-zone method and solid state reaction, respectively, as described elsewhere.\cite{lsvo2}
Polarized reflectivity spectra in the temperature range of 10-800 K were measured on the single crystals between 0.01 and 6 eV.
The reflectivity spectra between 5 and 40 eV were measured at room temperature
with use of synchrotron radiation at UV-SOR, Institute for Molecular Science.
For the Kramers-Kronig analysis, the spectrum above 40 eV was extrapolated by $\omega ^{-4}$ function,
while below 0.01 eV the Hagen-Rubens relation or the constant reflectivity was assumed
depending on the ground-state nature (metal or insulator).
$S$ was measured using a conventional steady state technique, in a cryostat (an electric furnace) below (above) room temperature.
A temperature gradient was generated by heating (cooling) one edge of the sample with a small resistive heater (an air pump).
The contribution of Au (Pt) wires for electrical contact was carefully subtracted.
While $x$ in the obtained single crystals is limited up to 0.26,
$S$ shows the nearly identical temperature and doping dependence between the single-crystalline and polycrystalline samples as shown later,
and thus we discuss the doping variation of $S$ mainly based on the results of the polycrystals prepared in a wider doping region.
All the measurements above room temperature were performed in a flow of Ar/H$_2$ (93\%/7\%) gas to prevent the oxidation
and we confirmed that the sample degradation at elevated temperature was negligible for the thermally-repeated measurements.

For the DMFT calculation in the single-band Hubbard model, the iterated perturbation theory was used to solve the effective impurity problem.
This solution is suitable for capturing general features of the correlation effects over a wide temperature.
Semi-circular density of states with the bandwidth of $2W$ was imposed on the non-interacting part of Hamiltonian.

\section{Results and discussion}

Figures 1 shows the temperature dependence of the resistivity in {\LSVO} up to high temperature.
The end compound LaVO$_3$ ($x=0$), a prototypical Mott insulator with the V electron configuration of $3d^2$,
undergoes the successive spin ordering (SO) and orbital ordering (OO) transitions
at $T_{\mathrm{SO}}=143$ K \cite{lsvomag1,lsvomag2,lsvomag3} and $T_{\mathrm{OO}}=141$ K.\cite{lsvo2,rvophase}
As shown in Fig. 1, the resistivity decreases dramatically with increasing the hole-doping level $x$,\cite{lsvores1,lsvores2}
and then shows the insulator-metal transition at $x_{\mathrm{IM}}=0.176$, accompanying the melting of the orbital order.\cite{lsvo2,lsvo3}
Above $x_{\mathrm{IM}}$, the resistivity shows a metallic behavior;
the difference observed between the single-crystalline and polycrystalline samples are due to grain boundary effects.
Above 200 K, the resistivity in the single crystals
asymptotically approaches the saturation values, indicating the incoherent charge transport.

\begin{figure}
\begin{center}
\includegraphics*[width=7.0cm]{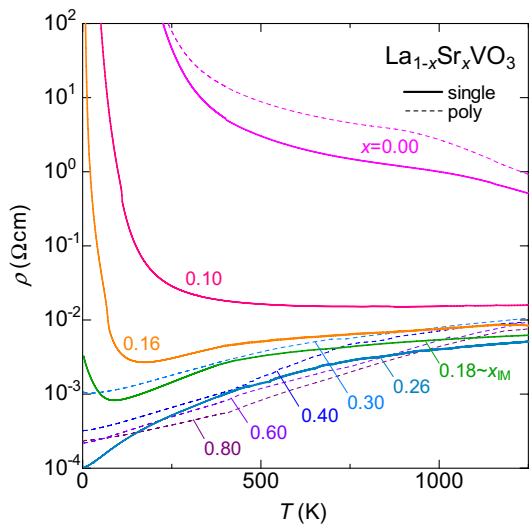}
\caption{
(Color online).
Temperature dependence of the resistivity $\rho$ in {\LSVO}.
The solid and dashed lines represent the data for the single-crystals and polycrystals, respectively.
}
\label{fig1}
\end{center}
\end{figure}

In Fig. 2(a), we show the doping variation of the optical conductivity spectra at $T=10$ K
for the light polarization parallel to the $c$ axis ({\EparaC}).
With increasing $x$, the spectral weight of the Mott-gap excitation around 2 eV decreases,
while the mid-infrared (mid-IR) peak appears in the inner-gap region and then evolves into the Drude peak at around $x_{\mathrm{IM}}$.\cite{lsvo3,lsvo4}
As shown in the Figs. 2(b) and (c),
the low-energy ($\hbar \omega<0.2$ eV) optical conductivity spectra near $x_{\mathrm{IM}}$
tends to show a broadened feature with increasing temperature, indicating the incoherent charge dynamics;
for $x=0.10$, the mid-IR peak evolves into the $\omega$-flat structure, accompanying the closing of the gap.
Also for $x=0.26$, the sharp Drude peak changes to the rather $\omega$-flat feature with increasing temperature.

\begin{figure}
\begin{center}
\includegraphics*[width=7.6cm]{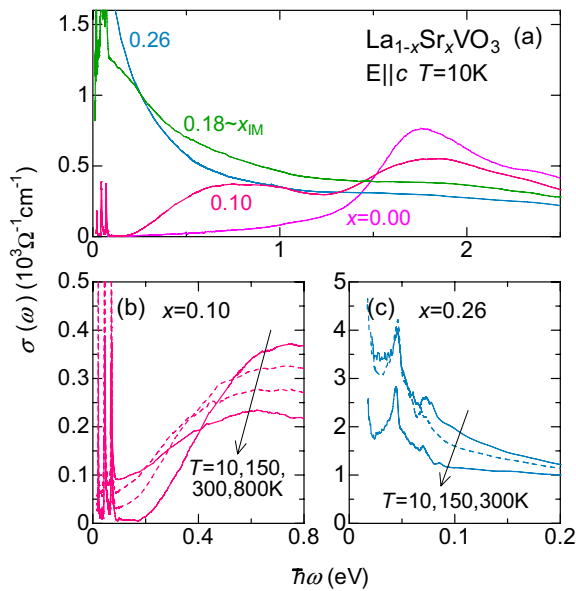}
\caption{
(Color online).
(a) Doping variation of the optical conductivity spectra at $T=10$ K
for the light polarization parallel to the $c$ axis ({\EparaC}).
Temperature dependence of the optical conductivity spectra for (b) $x=0.10$ and (c) 0.26 up to high temperature.
The $\omega$-flat features appear with increasing temperature, indicating the incoherent charge dynamics. 
}
\label{fig2}
\end{center}
\end{figure}

In Fig. 3, we show the temperature variation of $S$ in {\LSVO} up to 1250 K.
Below $x \simeq 0.1$, $S$ shows large positive values and decreases with increasing temperature,
being typical of insulators or semiconductors.
Above $x \simeq 0.4$, it shows linear $T$-dependence with a small slope, which represents the conventional metallic feature.
On the other hand, non-monotonic temperature dependence is observed centered around $x_{\mathrm{IM}}$;
the enhanced negative slope at low temperatures around $x_{\mathrm{IM}}$ indicates
the correlated metallic state with the enhanced carrier mass on the verge of the Mott transition,\cite{br}
which has also been confirmed by the measurement of the electronic specific heat coefficient in this system.\cite{lsvo2}
With increasing temperature, $S$ changes its sign at $T\sim$ 200 K,
and then reaches a relatively large positive value ($\gtrsim 20$ $\mu \mathrm{V/K}$) at $T\sim$ 600 K.
As described above, the results of the resistivity and the optical conductivity indicate that
the incoherent transport appears above 200 K around $x_{\mathrm{IM}}$.
Therefore, the change to the large positive values can be interpreted as the crossover to
the high-temperature regime where the Heikes formula holds valid.

\begin{figure}
\begin{center}
\includegraphics*[width=7.0cm]{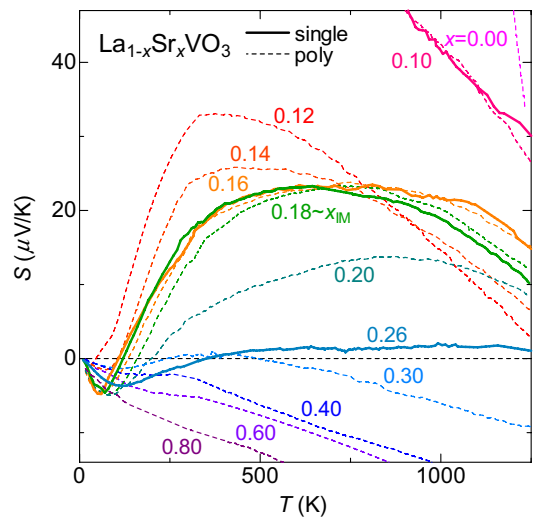}
\caption{
(Color online).
Temperature dependence of the thermopower $S$ in {\LSVO}.
The solid and dashed lines represent the data for the single-crystals and polycrystals, respectively.
}
\label{fig3}
\end{center}
\end{figure}

Figure 4(a) depicts the evolution of $S$ as the contour map in the $T-x$ plane.
The regions below $x \simeq 0.1$ and above $x \simeq 0.4$ indicate the prototypical insulating and band-metallic state, respectively.
In contrast, we can clearly see the region with large gradient of $S$ above 200 K centered around $x_{\mathrm{IM}}$, 
in which the incoherent charge dynamics is observed and the Heikes formula is expected to be applicable.

\begin{figure}
\begin{center}
\includegraphics*[width=8.0cm]{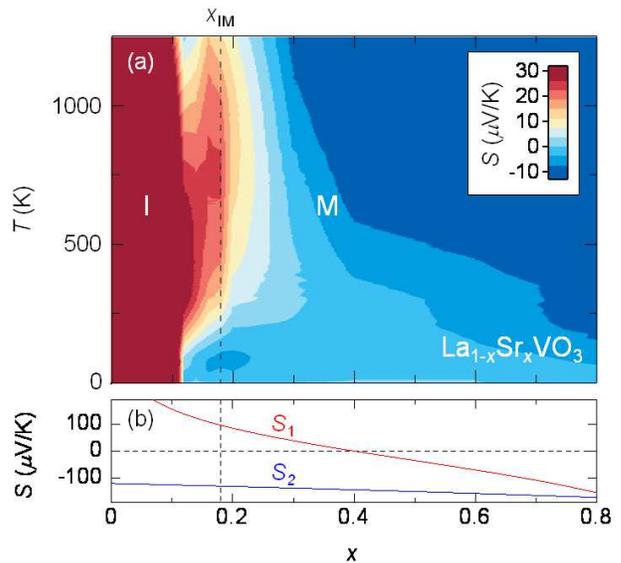}
\caption{
(Color online).
(a) Evolution of the thermopower $S$ in the $T$-$x$ phase diagram.
The vertical dashed lines indicate the critical doping level $x_{\mathrm{IM}}$
for the insulator (I)-metal (M) transition at the ground state.
The region with large $S$-gradient around $x_{\mathrm{IM}}$
corresponds to the incoherent metallic state as expected for the Heikes formula to be valid.
(b) Doping variation of $S$ in two high-temperature limits $S_1$ ($k_{\mathrm{B}}T \ll U$) and $S_2$ ($U \ll k_{\mathrm{B}}T$)
in the Heikes formula with consideration of nearly-degenerate $t_{2g}$ or ($t_{2g}$ plus $e_{g}$) orbitals (see text).
}
\label{fig4}
\end{center}
\end{figure}

The observed variation of $S$ around $x_{\mathrm{IM}}$ is explained
in terms of the two crossovers approaching the high-temperature limit values obtained from the Heikes formulas of Eqs. (1) and (2),
while $S$ is negative at low temperatures reflecting the sign of the charge carrier.
The theoretical asymptotic values in the high-temperature limits show the $x$ dependence, as shown in Fig. 4(b),
which are obtained by applying the Heikes formula to the {\LSVO} system.
In the case of (i) $k_{\mathrm{B}}T$ $\ll$ $U$,
the configurations are $g_e=9$ and $g_h=6$ for V$^{3+}$ (3$d^2$) and V$^{4+}$ (3$d^1$) ions, respectively,
and Eq. (1) is expressed as
\begin{equation}
\begin{split}
S_1=-\frac{k_{\mathrm{B}}}{e}\ln \frac{3x}{2(1-x)}.
\end{split}
\end{equation}
Here, we assumed that the crystal field splitting ($E_{\mathrm{c}}$) between the $t_{2g}$ and $e_{g}$ states
and the Hund's-rule coupling energy ($J_{\mathrm{H}}$) are all comparable to or smaller than $U$, but much larger than $k_{\mathrm{B}}T$.
For (ii) $U$, $E_{\mathrm{c}}$, $J_{\mathrm{H}}$ $\ll$ $k_{\mathrm{B}}T$, on the other hand, inserting $m=5$ and $n =2-x$ into Eq. (2), we have
\begin{equation}
\begin{split}
S_2=-\frac{k_{\mathrm{B}}}{e}\ln \frac{8+x}{2-x}.
\end{split}
\end{equation}
Around $x_{\mathrm{IM}}$, the limit values $S_1$ and $S_2$ are positive and negative, respectively.
Thus, the sign change at the characteristic temperature $T^*\sim$ 200 K
is due to the crossover asymptotically approaching $S_1$,
where the charge transport changes from the coherent motion in the correlated metal to the carrier hopping as in a doped Mott insulator.
Reflecting the doping variation of $S_1$, to which the spin and orbital degrees of freedom explicitly contribute,
the maximum value of $S$ in the intermediate temperature monotonically decreases with increasing $x$ up to $x \simeq 0.3$,
as shown in Fig. 3.
With further increasing temperature, $S$ begins to decrease while approaching negative $S_2$.
This may correspond to another crossover to the incoherent metal state allowing the doubly occupied states.
All these experimental results reveal that the energy scales $k_{\mathrm{B}}T$ ($\sim 20$ and 120 meV)
as corresponding to two crossover temperatures appear still smaller than the anticipated $W$ and $U$ values.
This should be considered not as discrepancy from the assumption in the derivation of $S_1$ and $S_2$ (Eqs. (3) and (4)),
but as reflection of the high sensitivity of $S$ to the entropy transport.
Such an \textit{undershooting} behavior of the $S$ crossovers is also partly reproduced in the simple DMFT calculation (\textit{vide infra}). 

Figure 5 shows the temperature dependence of the Hall coefficient $R_{\mathrm{H}}$ at $x=0.18$ $(\sim x_{\mathrm{IM}})$, as a typical example.
$R_{\mathrm{H}}$ still shows nearly-constant negative values reflecting the sign of the charge carrier,
while $S$ changes its sign from negative to positive.
The contrasting behavior between $S$ and $R_{\mathrm{H}}$ comes from the high sensitivity of $S$ to the entropy transport,
and indeed ensures that the $S$ crossovers cannot be explained by the change of the band structure around the Fermi level.
Thus, comparison with a model calculation which may be simple but fully takes into account the correlation effects will be useful
to capture the essence of the high-temperature $S$.

\begin{figure}
\begin{center}
\includegraphics*[width=6.8cm]{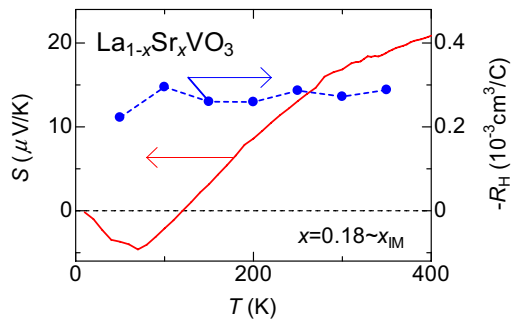}
\caption{
(Color online).
Comparison of temperature dependence between the thermopower $S$ and the Hall coefficient $R_{\mathrm{H}}$ at $x=0.18$.
}
\label{fig5}
\end{center}
\end{figure} 

The remarkable variation of $S$ in the Mott critical state, as typified by the sign changes of $S$ centered around $x_{\mathrm{IM}}$,
is shown using the selected experimental data set of Fig. 6(a).
Figures 6(b) and (c) show results of the DMFT calculation of $S$ with the doping rate $x$ and the ratio of $U/W$, respectively.
In this calculation, the energy gradient of density of states at Fermi level is positive at $k_{\mathrm{B}}T=0$,
and thus $S$ at low temperatures is negative since $-S$ is approximately proportional to the gradient.
In the high-temperature limit, on the other hand, $S$ must asymptotically approach the limit values 
given by
\begin{equation}
\begin{split}
S_1=-\frac{k_{\mathrm{B}}}{e}\ln \frac{2x}{1-x}
\end{split}
\end{equation}
for the case (i) $k_{\mathrm{B}}T$ $\ll$ $U$ ($g_e=2, g_h=1$) and 
\begin{equation}
\begin{split}
S_2=-\frac{k_{\mathrm{B}}}{e}\ln \frac{1+x}{1-x}
\end{split}
\end{equation}
for the case (ii) $U$ $\ll$ $k_{\mathrm{B}}T$ ($m=1, n=1-x$).
For $x=0.2$, as an example, the limit values $S_1$ and $S_2$ are positive and negative, respectively,
and thus $S$ should change its sign twice in the whole temperature region, as in the case of {\LSVO}.
As shown in Fig. 6(b), $S$ at $x=0.2$ indeed shows the non-monotonic temperature dependence analogous to the experimentally observed one.
With increasing $x$, $S_1$ decreases and finally changes the sign,
while the absolute value $|S_2|$ increases monotonically.
Besides, the Coulomb interaction becomes less effective for larger $x$ or in going away from the half-filling;
this explains the detailed $x$ dependence of the asymptotic behavior of $S$ (Fig. 6(b)).
Of course, except in the strong-correlation ($U \gg W$) and
high-temperature ($k_{\mathrm{B}}T \gg W$) limits,
$S$ is not in agreement with the limit values $S_1$ and $S_2$.
As shown in Fig. 6(c), in the case of $U/W=1.5$,
$S$ indeed shows the approach to $S_2$ before fully approaching $S_1$ ($\sim 60$ $\mu \mathrm{V/K}$).
With decreasing $U$, $S$ tends to directly approach $S_2$ ($\sim -35$ $\mu \mathrm{V/K}$),
while the carrier mass enhancement, manifesting as a steep negative $T$-gradient at low temperatures, disappears.
These theoretical results can reproduce the experimentally observed features for {\LSVO},
confirming that the effect of Coulomb interaction is essential for the observed non-monotonic temperature dependence of $S$. 

\begin{figure}
\begin{center}
\newpage
\includegraphics*[width=8.6cm]{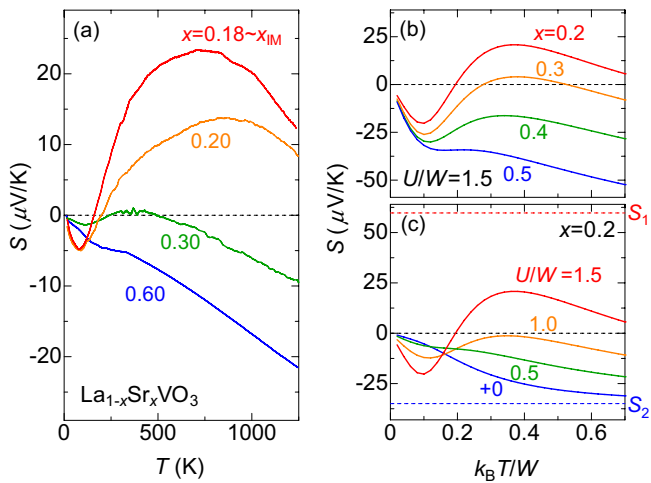}
\caption{
(Color online).
(a) Selected experimental data of the temperature and doping variation of the thermopower $S$ in {\LSVO}
for comparison with the theoretical results, (b) and (c).
Variation of $S$ with (b) doping $x$ and (c) Coulomb interaction $U$ obtained by the DMFT calculations on the single-band Hubbard model. 
}
\label{fig6}
\end{center}
\end{figure} 

\section{Conclusion}

Our systematic investigations on {\LSVO} thus show that
in the incoherent transport region above $T^* \sim 200$ K,
$S$ measures the entropy per charge carrier in response to the magnitude relation between $k_{\mathrm{B}}T$, $W$, and $U$ (and $E_{\mathrm{c}}$, $J_{\mathrm{H}}$),
and thus shows the two essential crossovers, asymptotically approaching respectively the limit values $S_1$ and $S_2$
obtained from the Heikes formulas.
In the case of some metallic layered cobaltites including {\NaxCo},
typical thermoelectric materials with the strong electron correlation,
the limit values have the same positive sign as the low-temperature one determined by the band structure,
and thus it would be difficult to detect the $S$ crossovers.
Recently, however, a similar crossover at $T^*$ ($\sim 200$ K) has been suggested for {\NaxCo}
by temperature-dependent photoemission spectroscopy; \cite{NaxCo6}
the unchanged sign of $S$ may be one of the origins of the large $S$ observed in these compounds. \cite{kotliar2}
The present findings unravel the generic features of the correlation effects on the high-temperature thermoelectric response
and demonstrate the possibility that
the Heikes formula can be indeed useful for the design of the correlated thermoelectric materials
at practical temperatures ($T>200$ K), not necessarily in the high-temperature limit.

\begin{acknowledgments}
We thank H. Sakai and Y. Ishida for helpful discussions.
This work was partially supported by Grants-in-Aid for Scientific Research (Grants No. 20046004 and No. 20340086) from the MEXT of Japan and JSPS
and Funding Program for World-Leading Innovative R \& D on Science and Technology (FIRST Program). 
M.U. acknowledges support by a Grant-in-Aid from the JSPS (Grant No. 21-5941).
\end{acknowledgments}


\begin{thebibliography}{100}
\bibitem{review1} G. Mahan, B. Sales, and J. Sharp, Phys. Today \textbf{50}(3), 42 (1997).
\bibitem{NaxCo0} I. Terasaki, Y. Sasago, and K. Uchinokura, Phys. Rev. B \textbf{56,} R12685 (1997).
\bibitem{NaxCo1} M. Lee, L. Viciu, L. Li, Y. Wang, M. L. Foo, S. Watauchi, R. A. Pascal, Jr., R. J. Cava, and N. P. Ong, Nat. Mater. \textbf{5,} 537 (2006).
\bibitem{koshimae1} W. Koshibae, K. Tsutsui, and S. Maekawa, Phys. Rev. B \textbf{62,} 6869 (2000).
\bibitem{koshimae2} W. Koshibae and S. Maekawa, Phys. Rev. Lett. \textbf{87,} 236603 (2001).
\bibitem{koshimae3} S. Maekawa, T. Tohyama, S. E. Barnes, S. Ishihara, W. Koshibae, and G. Khaliullin, \textit{Physics of Transition Metal Oxides} (Springer, Berlin, 2004).
\bibitem{NaxCo5} Y. Wang, N. S. Rogado, R. J. Cava, and N. P. Ong, Nature \textbf{423,} 425 (2003).
\bibitem{kotliar} G. P\'{a}lsson and G. Kotliar, Phys. Rev. Lett. \textbf{80,} 4775 (1998).
\bibitem{heikes1} R. R. Heikes and R. W. Ure, \textit{Thermoelectricity : Science and Engineering} (Interscience, New York, 1961).
\bibitem{heikes2} P. M. Chaikin and G. Beni, Phys. Rev. B \textbf{13,} 647 (1976).
\bibitem{NaxCo2} D. J. Singh, Phys. Rev. B \textbf{61,} 13397 (2000).
\bibitem{NaxCo3} T. Takeuchi, T. Kondo, T. Takami, H. Takahashi, H. Ikuta, U. Mizutani, K. Soda, R. Funahashi,
M. Shikano, M. Mikami, S. Tsuda, T. Yokoya, S. Shin, and T. Muro, Phys. Rev. B \textbf{69,} 125410 (2004).
\bibitem{NaxCo4} K. Kuroki and R. Arita, J. Phys. Soc. Jpn. \textbf{76,} 083707 (2007).
\bibitem{mott} N. F. Mott, \textit{Metal-Insulator Transitions} (Taylor and Francis, London, 1990).
\bibitem{MIT} M. Imada, A. Fujimori, and Y. Tokura, Rev. Mod. Phys. \textbf{70,} 1039 (1998).
\bibitem{lsvo1} F. Inaba, T. Arima, T. Ishikawa, T. Katsufuji, and Y. Tokura, Phys. Rev. B \textbf{52,} R2221 (1995).
\bibitem{lsvo2} S. Miyasaka, T. Okuda, and Y. Tokura, Phys. Rev. Lett. \textbf{85,} 5388 (2000).
\bibitem{lsvomag1} V. G. Zubkov, G. V. Bazuev, V. A. Perelyaev, and G. P. Shveiken, Sov. Phys. Solid State \textbf{15}, 1079 (1973).
\bibitem{lsvomag2} A. V. Mahajan, D. C. Johnston, D. R. Torgeson, and F. Borsa, Phys. Rev. B \textbf{46,} 10966 (1992).
\bibitem{lsvomag3} P. Bordet, C. Chaillout, M. Marezio, Q. Huang, A. Santoro, S. W. Cheong, H. Takagi, C. S. Oglesby, and B. Batlogg, J. Solid State Chem. \textbf{106,} 253 (1993).
\bibitem{rvophase} S. Miyasaka, Y. Okimoto, M. Iwama, and Y. Tokura, Phys. Rev. B \textbf{68,} 100406(R) (2003).
\bibitem{lsvores1} P. Dougier and P. Hagenmuller, J. Solid State Chem. \textbf{15,} 158 (1975).
\bibitem{lsvores2} A. V. Mahajan, D. C. Johnston, D. R. Torgeson, and F. Borsa, Phys. Rev. B \textbf{46,} 10973 (1992).
\bibitem{lsvo3} J. Fujioka, S. Miyasaka, and Y. Tokura, Phys. Rev. Lett. \textbf{97,} 196401 (2006).
\bibitem{lsvo4} S. Miyasaka, Y. Okimoto, and Y. Tokura, J. Phys. Soc. Jpn. \textbf{71,} 2086 (2002).
\bibitem{br} W. F. Brinkman and T. M. Rice, Phys. Rev. B \textbf{2,} 4302 (1970).
\bibitem{NaxCo6} Y. Ishida, H. Ohta, A. Fujimori, and H. Hosono, J. Phys. Soc. Jpn. \textbf{76,} 103709 (2007).
\bibitem{kotliar2} K. Haule and G. Kotliar, \textit{Properties and Applications of Thermoelectric Materials}, edited by V. Zlati\'{c} and A. C. Hewson (Springer, New York, 2009).
\newpage
\end{thebibliography}
\end{document}